\documentclass[11pt]{article}
\usepackage{color}

\usepackage{latexsym}  
\usepackage{amsmath}
\usepackage{amssymb}
\usepackage{graphicx}
\usepackage{slashed}
\usepackage{bm}
\usepackage{color}
\usepackage{braket}
\usepackage{ulem}

\newcommand{\be}{\begin{equation}}
\newcommand{\ee}{\end{equation}}
\newcommand{\bea}{\begin{eqnarray}}
\newcommand{\eea}{\end{eqnarray}}
\newcommand{\bref}[1]{(\ref{#1})}

\begin{document}
\begin{titlepage}
\begin{flushright}
\today
\end{flushright}

\begin{center}
{
Neutrinoless Double Beta Decay and $<\eta>$ Mechanism in the Left-Right Symmetric Model
}
\end{center}

\begin{center}

\vspace{0.1cm}

{Takeshi Fukuyama%
\footnote{E-mail: fukuyama@rcnp.osaka-u.ac.jp}}
and
{Toru Sato
\footnote{E-mail: tsato@rcnp.osaka-u.ac.jp}}


{\small \it Research Center for Nuclear Physics (RCNP),
Osaka University, \\Ibaraki, Osaka, 567-0047, Japan}\\


\end{center}

\begin{abstract}
 The neutrinoless double beta decay is studied in the framework of left-right symmetric model. The coexistence of left and right handed currents induces rather complicated interactions between the lepton and hadron sectors, called $<\lambda>$ mechanism and $<\eta>$ mechanism in addition to the conventional effective neutrino mass $<m_\nu>$ mechanism. In this letter, we study the possible magnification of $<\eta>$ mechanism and the relatively vanishing of $<\lambda>$ mechanism. The importance to survey $0\nu\beta\beta$ decay of different nuclei for specifying new physics beyond the Standard Model is also discussed.  
  
\end{abstract}
\end{titlepage}
\section{Introduction}

Neutrinoless double beta ($0\nu\beta\beta$) decay is one of the key probes for the new physics beyond the Standard Model (BSM physics). In this letter, we consider this process in the framework of left-right ($L-R$) symmetric model \cite{Pati, Mohapatra}, where the decay is concerned with the correlations between the $L$-handed light neutrinos and the $R$-handed heavy neutrinos.
$L-R$ symmetric model in the SO(10) grand unified theory appears in the intermediate stage \cite{Georgi, Fuku} which includes
\bea
SO(10) &\supset& SU(4)_{PS}\times SU(2)_L\times SU(2)_R \nonumber \\
&\supset&SU(3)_c\times SU(2)_L\times SU(2)_R\times U(1)_{B-L}
\eea
and is related with the wide varieties of BSM physics besides $0\nu\beta\beta$ decay, like baryo-genesis via lepto-genesis and dark matters etc.
There are two conditions to realize $0\nu\beta\beta$ decay in the context of this framework \cite{Takasugi}.\\
1. $\nu_e$ should be the same as its anti-particle  
\be
\nu_e=\overline{\nu_e}
\ee
and\\
2. the connecting neutrinos should have the same helicity.
The latter condition is satisfied if neutrinos are massive or if the $R$-handed current coexists with the $L$-handed current.
The first case of 2. is described as the well known effective neutrino mass,
\be
< m_\nu> =|\sum_j U_{ej}^2m_j|.
\label{effective}
\ee
Here $U_{\alpha i}$ (Greek (Latin) indicates flavour (mass) eigenstate) is the Pontecorvo-Maki-Nakagawa-Sakata (PMNS)
mixing matrix \cite{PMNS1, PMNS2} in $L$-handed current.
Substituting the observed values,
\be
|U_{11}|^2/|U_{13}|^2\approx 30.
\ee
Then, the inverted hierarchy (IH) case enhances $< m_\nu>$ relative to the normal hierarchy (NH) case.
Though the final answer to the hierarchy problem is given by observation, the theoretical predictions have been given by many models. One of the typical models is due to the predictive minimal SO(10) model \cite{Fuku}.
Based on this model, we fitted the low energy spectra of all quark and lepton masses and the CKM and the PMNS mixing angles and their phases.
Our results prefer the NH manifestly to the IH:  That is, inputting the observed lepton masses and the PMNS angles into the model, we compared the outputs of quark masses and CKM matrices \cite{Cabibbo, K_M} in the model with the observations. We obtained $\chi^2\leq 1$ for the NH case and $\chi^2>200$ for the IH case \cite{Fukuyama}. 
In this model, the effective neutrino mass is also predicted including the Majorana phases as
\be
< m_\nu> \approx 1~ \mbox{meV}.
\ee
On the other hand, the recent $0\nu\beta\beta$ experiment in the KamLAND-Zen \cite{KamLAND} provides the most stringent upper limit on it.  The half life $T_{1/2}$ of $0\nu\beta\beta$ decay in ${}^{126}$Xe is
\be
\frac{1}{T_{1/2}}=G_{0\nu}|M_{0\nu}|^2\left(\frac{<m_\nu>}{m_e}\right)^2<\frac{1}{2.3\times 10^{26} ~\mbox{yr}}~ @ ~90\%~ C.L.
\label{T1/2a}
\ee
Here $G_{0\nu}$ is the phase-space integral and $M_{0\nu}$ is the nuclear matrix element (NME), which leads to the $<m_\nu>=36-156$ meV \cite{KamLAND, GERDA, CUORE}, already in the inverted hierarchy regions. 
Efforts to reduce the ambiguities of NME in different nuclear models
are in progress~\cite{suhonen98,Engel,Yao,cirigliano22}.

  We consider $0\nu\beta\beta$ decay in the $L-R$ symmetric model.
   The model generates contact, heavy neutrino
  and light neutrino exchange quark-quark interactions
  The quark-quark interactions then mapped onto contact, pion exchange and
  light neutrino exchange interactions between two nucleons
  classified in a general from in effective field
  theory~\cite{Cirigliano1,Cirigliano2}.
  Here we focus on the light neutrino exchange mechanism
  in particular the $<\lambda>$ and the $<\eta>$ mechanisms in addition
  to the $<m_\nu>$ mechanism~\cite{Takasugi,Tomoda}.
  The BSM physics appears in the leptonic currents, which restricts the
  structure of the hadronic current. 
  Such interplay between leptonic and hadronic
  currents has been overlooked so far.
  Especially, $0\nu\beta\beta$ is very sensitive to the spatial momentum of neutrino propagator and
  interference between the vector and axial vector currents of nucleon,
  enhancing $<\eta>$ mechanism~\cite{Takasugi,tomoda85,stefanik}.
  This term is also very crucial to the heavy right-handed neutrino mass.
  In this letter, surveying into these entanglements of interplay and
  solving them, we narrow down the general forms of
  L-R symmetric models~\cite{Cirigliano1, Cirigliano2}, leading to the
  mechanism of $0\nu\beta\beta$ to $<m_\nu>$ and $<\eta>$ mechanisms
 if experiments reveal the non-null result around the present upper bound.
 
 This letter is organized as follows. In Sec.2,  we discuss the structure of the leptonic current, assuring the low energy seesaw mechanism. Its hadronic counterparts are studied in Sec.3.
 We show a simple understanding of the mechanism to enhance
 a sensitivity to the $<\eta>$ mechanism due to the V-A interference term.
   This mechanism has a potential to reveal $R$-handed current
   if the  $0\nu\beta\beta$ decay is discovered above the NH region from the
   present and near future experiments.
   It is also discussed that the atom dependence (A-dependence) of $0\nu\beta\beta$ beta decay rate
   may clarify  $<m_\nu>$ and/or $<\eta>$ mechanisms.
Section 4 is devoted to discussions.
 
 \section{Right-handed  weak current}
 We consider $L-R$ symmetric model \cite{Mohapatra} in this section. The weak Hamiltonian is given by
 \be
H_W=\frac{G_F\cos\theta_c}{\sqrt{2}}\left[j_L^\mu \tilde{J}_{L\mu}^\dagger+
   j_R^\mu \tilde{J}_{R\mu}^\dagger\right]+H.c.
\label{Fermi}
\ee

Here $j_{\mu}$ ($J_{\mu}$) indicates leptonic (hadronic) current, and the  $L$ and $R$-handed leptonic currents, $j_{L\mu}$ and $j_{R\mu}$, are given by
\bea
\label{Leptonic1}
j_{L\alpha}&=& \sum\overline{l(x)}\gamma_\alpha 2P_L\nu_{lL}(x),\\
j_{R\alpha}&=& \sum\overline{l(x)}\gamma_\alpha 2P_RN_{lR}(x),
\label{Leptonic2}
\eea
where
\be
P_L\equiv \frac{1}{2}(1-\gamma_5),~~P_R\equiv\frac{1}{2}(1+\gamma_5).
\ee
 Also $\nu_{lL}(N_{lR})$ are $L$-handed ($R$-handed) weak eigenstates of the neutrinos, and
\bea
\label{Hadronic1}
  \tilde{J}_L^\mu(\bm{x}) & = & J_L^\mu(\bm{x}) + \kappa J_R^\mu(\bm{x}),
  \\
  \tilde{J}_R^\mu(\bm{x}) & = & \eta J_L^\mu(\bm{x}) + \lambda J_R^\mu(\bm{x}).
  \label{Hadronic2}
  \eea
  Thus the system is mixed with rather simple leptonic world and composite hadronic world. There are many precedent works to have discussed $0\nu\beta\beta$ in the $L-R$ symmetric models \cite{Awasthi, Dev, Deppisch, Li, Boruah} etc.
  We are not concerned with the detailed new calculations of hadronic models but to try to give a firm foundation for low energy seesaw mechanism and to make clear the connection of neutrino potential with hadronic NMEs. The main diagrams of 0$\nu\beta\beta$ decay in the $L-R$ symmetric model are depicted in Fig.1. 
\begin{figure}
\begin{center}
  \includegraphics[width=15cm]{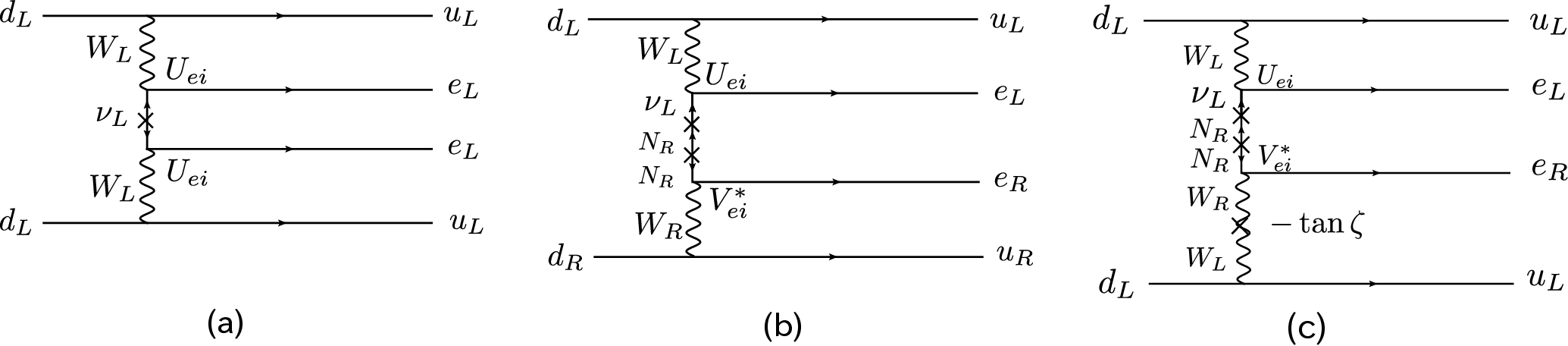}
 \caption{Diagrams of $0\nu\beta\beta$ decay. (a), (b), and (c) are $< m_\nu>,~< \lambda>$, and $<\eta>$-mechanisms, respectively.}
\end{center}
\end{figure}  
\begin{figure}
\begin{center}
  \includegraphics[width=6cm]{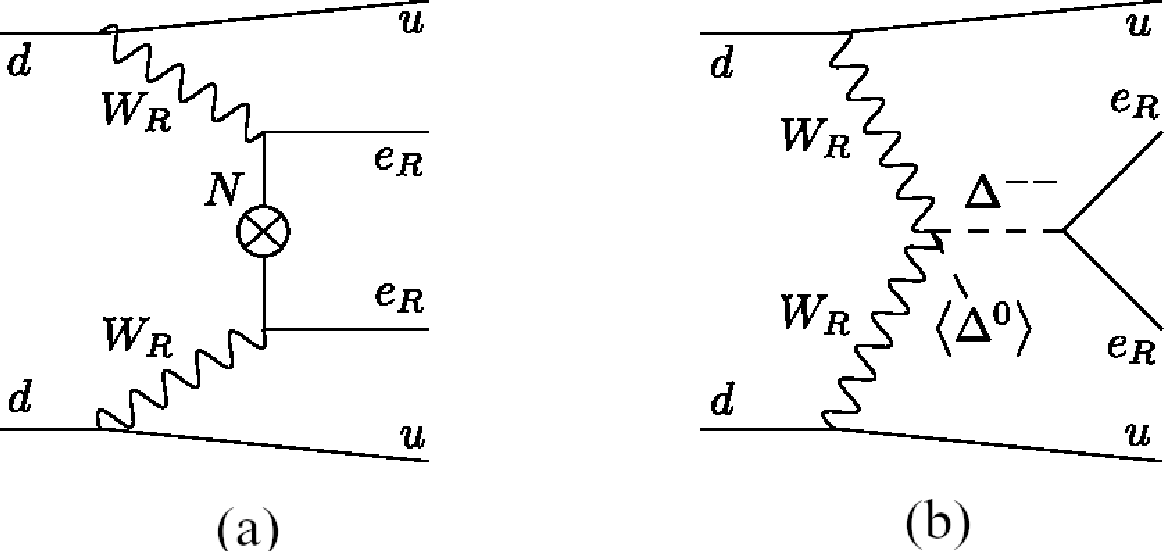}
 \caption{Diagrams of $0\nu\beta\beta$ decay via $W_R-W_R$ lines. }
\end{center}
\end{figure} 
In general, we have the other diagrams via $W_R-W_R$ lines Fig.2. However,  $W_R-W_R$ contributions give the result as \cite{Fuku2}
\bea
\frac{A_R^{(a)}}{A_R^0}&=&0.15\times \frac{g_R^4}{g_L^4}\left(\frac{5\mbox{TeV}}{M_{WR}}\right)^4\frac{100\mbox{GeV}}{m_N}\\
\frac{A_R^{(b)}}{A_R^0}&=&0.15\times \frac{g_R^4}{g_L^4}\left(\frac{5\mbox{TeV}}{M_{WR}}\right)^4\frac{<\Delta^0>}{8\mbox{TeV}}\left(\frac{1\mbox{TeV}}{m_{\Delta^{++}}}\right)^2\frac{g_{ee}}{0.3}.
\eea
Here $A_R^0$ is the current experimental bound. $A_R^{(a),(b)}$ are the amplitude of (a) and (b) of Fig.2.  Thus these diagrams give subdominant contributions. So we limit our arguments in Fig.1 hereafter. 
Its amplitude in closure approximation \cite{Tomoda}
  is given as
\be
R_{0\nu}=4\sqrt{\frac{1}{2}}\left(\frac{G\cos\theta_c}{\sqrt{2}}\right)^2
\sum_i\sum_{\alpha,\beta}\int d\bm{x}d\bm{y}\int\frac{d\bm{k}}{(2\pi)^3}
e^{i\bm{k}\cdot (\bm{y}-\bm{x})} H^{\nu\mu}  L_{\nu\mu},
\label{R0nu1}
\ee
where the lepton tensor $L^{\nu\mu}$ is
\be
L_{\nu\mu}=\overline{e}_{p_2,s_2'}(\bm{y})\gamma_\nu P_\beta\frac{1}{2\omega}
\left[\frac{\omega\gamma^0-\bm{k}\cdot\bm{\gamma}+m_i}{\omega+A_1}
  + \frac{-\omega\gamma^0 -\bm{k}\cdot\bm{\gamma}+m_i}{\omega+A_2}
  \right]P_\alpha\gamma_\mu e_{p_1,s_1'}^c(\bm{x}).
\label{eq:lmunu}
\ee
Here $e_{p_i,s_i'}(\bm{x})$ are electron wave functions with the energy $e_i$, and  the mixing matrices $U,~V^*$ are omitted for simplicity.
The energy denominator is given by $A_i = e_i + <E_n> - E_i$
and $E_f +e_1 + e_2 = E_i$.
Here $E_{i/f}$ and $<E_n>$ are energy of the initial/final nuclear state
and the average  energy of the intermediate nuclear state, respectively. 

The nuclear tensor $H^{\nu\mu}$ is given by the matrix element of the nuclear weak current as
\be
H^{\nu\mu} =
\braket{F|\tilde{J}_{\beta i}^{\nu +}(\bm{y})\tilde{J}_{\alpha i}^{\mu +} (\bm{x})|I},
\ee
where $\tilde{J}^\mu_{L,R}$ are given in \bref{Hadronic1} and \bref{Hadronic2}.

The neutrino propagator becomes,
\be
P_\alpha (\pm \omega\gamma^0  - \bm{k}\cdot\bm{\gamma} +m_i)P_\beta = \begin{cases}
m_iP_\alpha & (\alpha=\beta) \\
(\pm \omega\gamma^0  - \bm{k}\cdot\bm{\gamma}) P_\beta & (\alpha\neq \beta)
\end{cases}.
\label{L-R}
\ee
In the presence of the  R-handed current, 
we have $(\pm \omega\gamma^0  - \bm{k}\cdot\bm{\gamma})P_\beta$ in addition to
\bref{effective}. The spatial momentum exchanged between nucleon by neutrino
is significantly larger than neutrino mass term,
$|{\bf k}|\approx 100\mbox{MeV}\gg E_n-E_i, m_i$,
which gives a significant effect to the decay rate.
This mechanism gives interesting interplay between particle physics and
nuclear physics, whose explanation is the main theme of this paper. 

The half life $T_{1/2}$ in this case
\cite{Takasugi} is given as 
\begin{eqnarray}
   \frac{1}{T_{1/2}}& =&
  C_{mm}^{(0)} (\frac{<m_\nu>}{m_e})^2
  + C_{m\lambda}^{(0)}   \frac{<m_\nu>}{m_e}<\lambda>\cos\psi
  + C_{m\eta}^{(0)}   \frac{<m_\nu>}{m_e}<\eta>\cos\psi
  \nonumber \\
& &
  + C_{\lambda\lambda}^{(0)} <\lambda>^2
      + C_{\eta\eta}^{(0)} <\eta>^2
      + C_{\lambda\eta}^{(0)} <\lambda><\eta>.
\label{T1/2}
\end{eqnarray}
Here $C_{ab}^{(0)}$ includes NME and phase space integral. The other parts include BSM physics. The effective couplings
$<\eta>$ and $<\lambda>$ are given as
\be
<\lambda>=\lambda |\sum _j{}'  U_{ej}V_{ej}^*|,~~<\eta>=\eta|\sum_j  {}' U_{ej}V_{ej}^*|.
\label{LambdaEta}
\ee
$\psi$ is the relative phase between $<m_\nu>$ and $<\lambda>$ and $<\eta>$, 
\be
\psi=\mbox{arg}\left[\left(\sum{}'m_jU_{ej}^2\right)\left(\sum{}'U_{ej}V_{ej}^*\right)^*\right],
\label{psi}
\ee
where $\sum '$ indicates the summation over only the light neutrinos. However, $U$ and $V$ are independent and we set $\psi=0$ hereafter. The details of $\lambda$ and $\eta$ are given by \bref{lambda} and \bref{eta}.

We proceed to discuss the detailed structure of mixing matrices.
Higgs sectors are composed of  (3,1,2), (1,3,2) triplets ($\Delta_L$, $\Delta_R$, respectively) and bi-doublet (2,2,0) ($\Phi$) under $SU(2)_L\otimes SU(2)_R\otimes U(1)_{B-L}$. They have the vacuum expectation values $v_u,~v_d,~v_L,~v_R$ as
\be
<\Phi>_0=<\left(
\begin{array}{cc}
\phi_1^0&\phi_2^+\\
\phi_1^-&\phi_2^0\\
\end{array}
\right)>_0=\left(
\begin{array}{cc}
v_u&0\\
0&v_d\\
\end{array}
\right)
\label{vev1}
\ee
and 
\be
<\Delta_{L,R}>_0=<\left(
\begin{array}{cc}
\frac{\delta^+}{\sqrt{2}}&\Delta^{++}\\
\Delta^0&-\frac{\delta^+}{\sqrt{2}}\\
\end{array}
\right)_{L,R}>_0
=\left(
\begin{array}{cc}
0&0\\
v_{L,R}&0\\
\end{array}
\right).
\label{vev2}
\ee
The neutrino mass matrix is \cite{Minkowski, Yanagida, Gellmann, M_S}
\be
M_\nu=\left(
\begin{array}{cc}
M_L&M_D^T\\
M_D&M_R\\
\end{array}
\right)\approx
\left(
\begin{array}{cc}
0&M_D^T\\
M_D&M_R\\
\end{array}
\right).
\label{Mnu}
\ee
Thus we have the extended Fermi couplings \bref{Fermi}. In \bref{Leptonic1} and \bref{Leptonic2}, $\nu_{lL}(N_{lR})$ are $L$-handed ($R$-handed) weak eigenstates of the neutrinos. Using $3\times 3$ blocks $U,V,X,Y$, the mass eigenstates $\nu',~N'$ are given as
\be
\left(
\begin{array}{c}
\nu\\
(N_R)^c\\
\end{array}
\right)_L=\left(
\begin{array}{cc}
U&X\\
V&Y\\
\end{array}
\right)\left(
\begin{array}{c}
\nu'\\
N'\\
\end{array}
\right)_L
\equiv \mathcal{U}\left(
\begin{array}{c}
\nu'\\
N'\\
\end{array}
\right)_L.
\label{mixing}
\ee
That is,
\be
(\nu_L)_\alpha=U_{\alpha i}\nu_i'+X_{\alpha I}N_I',~~(N_R)^c_\alpha=V_{\alpha i}\nu_i'+Y_{\alpha I}N_I',
\ee
where $\alpha$ ($i$) are the flavour (mass) eigenstates.

The constants $\lambda$ and $\eta$  in \bref{Fermi} are related to the mass eigenvalues of the weak bosons in the $L$ and $R-$ handed gauge sectors ($W_L,~W_R$) as follows:
\bea
W_L&=&W_1\cos\zeta+W_2\sin\zeta,\\
W_R&=&-W_1\sin\zeta+W_2\cos\zeta,\\
\frac{G_F}{\sqrt{2}}&=&\frac{g^2}{8}\cos^2\zeta\frac{M_{W1}^2\tan^2\zeta+M_{W2}^2}{M_{W1}^2M_{W2}^2}, \\
\label{lambda}
\lambda&\equiv &\frac{M_{W1}^2+M_{W2}^2\tan^2\zeta}{M_{W1}^2\tan^2\zeta+M_{W2}^2},\\
\eta&\equiv & -\frac{(M_{W2}^2-M_{W1}^2)\tan\zeta}{M_{W1}^2\tan^2\zeta+M_{W2}^2}.
\label{eta}
\eea
Here $M_{W1}$ and $M_{W2}$ are the masses of the mass eigenstates $W_1$ and $W_2$, respectively, and $\zeta$ is the mixing angle which relates the mass eigenstates and the gauge eigenstates.
We are considering $L-R$ symmetric model. The gauge boson mass is generated from \bref{vev1} and \bref{vev2} as \cite{Zhang}
\be
M_W^2=\left(
\begin{array}{cc}
\frac{1}{2}g^2(v_u^2+v_d^2+2v_L^2)&g^2v_uv_d\\
g^2v_uv_d&\frac{1}{2}g^2(v_u^2+v_d^2+2v_R^2)\\
\end{array}
\right)
\ee
and the mixing angle $\zeta$ is
\be
\tan 2\zeta=\frac{2v_uv_d}{v_R^2-v_L^2}\approx \frac{2v_uv_d}{v_R^2}=2\xi\left(\frac{M_{WL}}{M_{WR}}\right)^2
\label{zeta}
\ee
with
\be
v_u^2+v_d^2=v_{ew}^2,
\ee
\be
\xi=v_d/v_u=1/\tan\beta
\ee
and
\be
M_{W2}=\sqrt{2}g_Rv_R\geq 5 \mbox{TeV}.
\ee
\cite{CMS, ATLAS}.
 In the $L-R$ symmetric model, we set $g_L=g_R$, which indicates further unification of at least rank five GUT, including SU(3) color. $\tan\beta$ is constrained from the Yukawa coupling is renormalizable up to the GUT scale,
\be
1\leq \tan\beta\leq 60.
\label{tanbeta}
\ee
That is, the upper limit (lower limit) appears from the renormalizability of bottom (top) Yukawa coupling in GUT. 
 Furthermore, in this case, large $\tan\beta$ induces too rapid proton decay since the proton life-time is proportional to $1/\tan\beta^2$ and $\tan\beta$ is limited around $10$ \cite{Fukuyama}. 
Reflecting these relatively low mass constraint, we will discuss on the low energy seesaw mechanism later.

Corresponding to Figure 1, we will consider $0\nu\beta\beta$ decay in this scheme:

\begin{itemize}
\item $W_L-W_L$ diagram
\be
m_{eff}^{LL}=\sum_{i=1}^3 U_{ei}^2m_i+\sum_{i=1}^{3,6}k^2X_{ei}^2\frac{M_I}{k^2-M_I^2}.
\label{LL}
\ee
Here and in the subsequent discussions in this section, we write the subdominant terms in addition to \bref{LambdaEta}, illustrating  the seesaw structure. In the latter sum, $3$ and $6$ indicate type I \bref{mixing} and Inverse seesaw mechanisms \bref{U3}, respectively.
\item $W_R-W_R$ diagram
\be
m_{eff}^{RR}=\sum k^2Y_{ei^*}^2\frac{M_I}{k^2-M_I^2}\frac{g_R^4}{g_L^4}\frac{M_{WL}^4}{M_{WR}^4},
\ee
with $g_L=g_R$, which was suppressed compared with the others due to $\left(\frac{M_{WL}}{M_{WR}}\right)^4$.
\item $W_L-W_R$ diagram; the neutrino mixing $(\lambda)$ and $W_L-W_R$ mixing $(\eta)$
\bea
<\lambda>&=&\left(U_{ei}V_{ei}^*+X_{ei}Y_{ei}^*\frac{k^2}{k^2-M_I^2}\right)\frac{M_{WL}^2}{M_{WR}^2},\\
<\eta>&=&\left(U_{ei}V_{ei}^*+X_{ei}Y_{ei}^*\frac{k^2}{k^2-M_I^2}\right)(-\tan \zeta).
\label{LR}
\eea
Here the first terms $U_{ei}V_{ei}^*$ dominate, and let us estimate the magnitude of $V_{ei}$. The naive type I seesaw \bref{Mnu} gives tiny value for the above quantities. We are interested in TeV scale seesaw and consider  the inverse seesaw mechanism \cite{Mohapatra2} hereafter. Its $9\times 9$ mass matrix is given by
\be
M_\nu=\left(
\begin{array}{ccc}
0&M_D^T&0\\
M_D&0&M^T\\
0&M&\mu
\end{array}
\right)\equiv
\left(
\begin{array}{cc}
0_{3\times 3}&\mathcal{M}_{D3\times 6}^T\\
\mathcal{M}_{D6\times 3}&\mathcal{M}_{R6\times 6}
\end{array}
\right).
\ee
 Their mass scales are $\mu\approx O(1) keV, ~M_D\approx O(100)$ GeV, ~$M\approx O(1)$ TeV \footnote{Such a low mass $M_R$ in SO(10) GUT is realized, for instance, by the radiatively generated $M_R$ model \cite{Witten}.}.
 This matrix is approximately diagonalized by $9\times 9$ mixing matrix $\mathcal{U}$ \cite{Awasthi, Barry, Paulo}
\bea
\label{U1}
&&\mathcal{U}\approx \\
&&\left(
 \begin{array}{cc}
 1-\frac{1}{2}\mathcal{M_D}^\dagger[\mathcal{M_R}(\mathcal{M_R})^\dagger]^{-1}\mathcal{M_D}&\mathcal{M_D}^\dagger(\mathcal{M_R}^\dagger)^{-1}\nonumber\\
 - \mathcal{M_R}^{-1}\mathcal{M_D}&1-\frac{1}{2}\mathcal{M_R}^{-1}\mathcal{M_D}\mathcal{M_D}^\dagger(\mathcal{M_R}^\dagger)^{-1}
 \end{array}
 \right)
 \eea
 as
 \be
 \mathcal{U}^TM_\nu\mathcal{U}=\left(
 \begin{array}{cc}
m_{light}&0_{3\times 6}\\
0_{6\times 3}&M_{heavy~ 6\times 6}
\end{array}
\right).
\label{U2}
\ee
Here
\be
m_{light}=M_D^TM^{-1}\mu (M^T)^{-1}M_D
\ee
and
\be
M_{heavy}=\mathcal{M}_R.
\ee
Thus $\mathcal{U}$ of \bref{mixing} in type I seesaw is modified in the inverse seesaw as
\be
\mathcal{U}=\left(
\begin{array}{cc}
U & X\\
V & Y\\
W & Z\\
\end{array}
\right),
\label{U3}
\ee
where $\{U,~V,~W\}$ and $\{X,~Y,~Z\}$ are $3\times 3$ and $3\times 6$ matrices, respectively.
All deviation from unitarity is determined by
\be
\zeta=\mathcal{M}_R^{-1}\mathcal{M}_D.
\ee
It goes from \bref{U2} and \bref{U3},
\be
\left(
\begin{array}{c}
V\\
W\\
\end{array}
\right)=-\mathcal{M}_R^{-1}\mathcal{M}_D=-\left(
\begin{array}{cc}
0&M\\
M&\mu\\
\end{array}
\right)^{-1}\left(
\begin{array}{c}
M_D\\
0\\
\end{array}
\right)=\frac{1}{M^2}
\left(
\begin{array}{c}
-\mu M_D\\
M M_D\\
\end{array}
\right).
\ee
Then
\be
V\approx O(\frac{\mu M_D}{M^2})=O\left(\frac{m_{light}}{M_D}\right)\ll O(\zeta)
\label{order}
\ee
and $O\left(\frac{m_{light}}{M_D}\right)$ is also valid for type I seesaw. Thus, $O(V)$ seems to be tiny. However, this is too naive estimation.
This is because these estimations are due to the hierarchy assumption $M\gg M_D\gg \mu$ and to the neglect of the generation. Alternative ideas may be free from the light neutrino mass constraint like \bref{order}. For type I seesaw, for example, if $(M_D)_{i3}=0$ and $(M_N^{-1})_{33}=0$, then, $m_\nu=0$ and $V$ is free from the light neutrino mass constraint \footnote{We are grateful to Mimura on this indication.}. For the case of inverse seesaw, we may consider another mass hierarchy $M\gg \mu\gg M_D$. Important is that there are windows of sizable $V$.
The order of the magnitude of $V$ should be estimated from the observations. 

As was shown in \bref{tanbeta}, $<\eta >$ is not so suppressed and its contribution in \bref{T1/2} may be important when we take into account the large contribution from NME $C_{\eta\eta}^{(0)}$ .

\end{itemize}

\section{Nuclear matrix elements and role of $<\eta>$ mechanism}
  The sensitivity of the  0$\nu\beta\beta$ decay to the R-handed
  current can be roughly summarized
  as follows. The decay rate calculated from the neutrino mass terms 
  with $<m_\nu> \sim 0.1$ eV corresponds to  the R-handed current
  contribution of either  $<\eta> \sim 10^{-9}$ or
  $<\lambda> \sim 10^{-7}$~\cite{suhonen98}.
  The  two order of magnitude 
  difference of the sensitivity between the $<\eta>$ mechanism and the $<\lambda>$
  mechanism
 comes from the 
 interference  between the nuclear vector and axial vector current.
 The combination of the $V-A$ interference of nuclear current, 
 which corresponds to NME $\chi_R$ and $\chi_P$~\cite{Takasugi},
 and the phase space integral $G_i$
 enhances sensitivity to the $<\eta>$ mechanism.
  In the previous section, it is argued that $<\eta>$ may not be
as tiny as usually assumed, which opens an
  interesting possibility to
  reveal R-handed current through the $<\eta>$ mechanism.
We revisit this
mechanism,  showing a
simplified derivation of the relevant nuclear operator
on the  $0^+-0^+$ 0$\nu\beta\beta$ decay.
 We then estimate the allowed region of $<\eta>$ from the current upper limit
 of the 0$\nu\beta\beta$ decay rare.

The transition amplitude $R_{0\nu}$ \bref{R0nu1}
can be written as
\begin{eqnarray}
R_{0\nu}&=&4\sqrt{\frac{1}{2}}\left(\frac{G\cos\theta_c}{\sqrt{2}}\right)^2\sum_i\sum_{\alpha,\beta}\int d\bm{x}d\bm{y}\int\frac{d\bm{k}}{(2\pi)^3}
e^{i\bm{k}\cdot (\bm{y}-\bm{x})}
\frac{1}{2\omega}\left[
  \frac{1}{\omega+A_1} + \frac{1}{\omega+A_2}\right]
\nonumber \\
& & \times \overline{e}_{p_2,s_2'}(\bm{y})
          \bra{F}{\cal O}(\bm{x},\bm{y}) \ket{I}e_{p_1,s_1'}^c(\bm{x}),
\label{R0nu2}
\end{eqnarray}
where ${\cal O}(\bm{x},\bm{y})$ is the NME of the hadronic current.
The interference terms of $R$-handed and $L$-handed current from the 
neutrino momentum ($\bm{k}$) dependent term of the neutrino propagator
\bref{L-R} are given as
\begin{eqnarray}
  {\cal O}(\bm{x},\bm{y}) & = &-
    \slashed{\tilde{J}}_R^\dagger(\bm{y})\bm{k}\cdot\bm{\gamma}P_L
    \slashed{\tilde{J}}_L^\dagger(\bm{x})
    -
    \slashed{\tilde{J}}_L^\dagger(\bm{y})\bm{k}\cdot\bm{\gamma}P_R
    \slashed{\tilde{J}}_R^\dagger(\bm{x}). \label{eq:oeff1}
\end{eqnarray}
The Dirac matrices are for the electron spinor.
Using the hadronic currents of \bref{Hadronic1} and \bref{Hadronic2},
we keep the interference terms between  the SM hadronic $L$-handed current
and the BSM  $L$-handed current $\eta J^\mu_L$ and
the $R$-handed current $\lambda J^\mu_R$, while
the $\kappa J_R^\mu$ term can be neglected.
Using  $J^\mu_{L/R} = V^\mu \pm A^\mu$,
  we keep only interference term between the vector and the
  axial vector current in ${\cal O}(\bm{x},\bm{y})$, 
\begin{eqnarray}
  {\cal O}(\bm{x},\bm{y}) & = &
 [
\slashed{V}(\bm{y}) \bm{k}\cdot\bm{\gamma}  \slashed{A}(\bm{x})
    \mp
\slashed{A}(\bm{y})  \bm{k}\cdot\bm{\gamma} \slashed{V}(\bm{x})] 
    \left ( \begin{array}{c}
    \gamma_5 <\lambda> \\
     <\eta>
  \end{array}
  \right )
\end{eqnarray}
Here $\mp$ is for $<\lambda>$ and $<\eta>$ term, respectively.
We keep the contribution of the spatial component of the axial vector current,
i. e. Gamow-Teller operator, as 
\begin{eqnarray}
{\cal O}(\bm{x},\bm{y}) & \sim &
  [ -i(V_0(\bm{y}) \bm{A}(\bm{x}) \pm \bm{A}(\bm{y})V_0(\bm{x}))\times \bm{k}
  \cdot\bm{\gamma}
  + i(\bm{V}(\bm{y}) \times \bm{A}(\bm{x}) \pm \bm{A}(\bm{y})\times \bm{V}(\bm{x}))
  \cdot \bm{k}\gamma_0 ] \nonumber \\
& & \times
  \left ( \begin{array}{c}
    <\lambda> \\
    \gamma_5 <\eta>
  \end{array}
  \right )
\end{eqnarray}
Here $\pm$ is for $<\lambda>$ and $<\eta>$ term, respectively.

The $V-A$ interference terms remains only for the $<\eta>$ mechanism.
The interference terms between the time component of the vector current($V_0$)
and the space component of the axial vector current($\bm{A}$)
including s-wave and p-wave electron contribute to the  $\chi_P$
term. The spatial components of the vector current($\bm{V}$)
and the axial vector current($\bm{A}$) with the s-wave electron contribute
to the $\chi_R$ term.

We examine further the $\bm{V}-\bm{A}$ term.
Either emitted electrons or nuclear currents have to compensate
the p-wave nature of neutrino propagator.
The main contribution comes from the s-wave electron
and momentum dependent nuclear magnetization current.
${\cal O}(\bm{x},\bm{y})$, which become scalar effective nuclear
operator for the $0^+-0^+$ transition, is given as
\begin{eqnarray}
  {\cal O}(\bm{x},\bm{y}) & = &
< \lambda > (
  \bm{k}\times \bm{\mu}(\bm{y})\cdot\bm{k}\times\bm{A}(\bm{x})
 -\bm{k}\times \bm{A}(\bm{y})\cdot\bm{k}\times\bm{\mu}(\bm{x})
  )(-\gamma_0)
    \nonumber \\ & &
    \ \ \ + < \eta > (
  \bm{k}\times \bm{\mu}(\bm{y})\cdot\bm{k}\times\bm{A}(\bm{x})
 + \bm{k}\times \bm{A}(\bm{y})\cdot\bm{k}\times\bm{\mu}(\bm{x})
  )(\gamma_5 \gamma_0).
\end{eqnarray}
Here the magnetization current of the vector current
is expressed as $\bm{V}(\bm{x}) = \bm{\nabla}\times \bm{\mu}(\bm{x})$.
In the non-relativistic and impulse approximation of nuclear current,
$\bm{\mu}(\bm{x})$ and $\bm{A}(\bm{x})$ are
given by  using the same spin-isospin flip 
operator $\sim \tau^+ \bm{\sigma} $ as~\cite{koshigiri79},
\begin{eqnarray}
  \bm{A}(\bm{x}) & = & \sum_i^A g_A(k^2) \tau^+_i \bm{\sigma}_i \delta(\bm{x} - \bm{r}_i), \\
  \bm{\mu}(\bm{x}) & = & \sum_i^A \frac{g_V(k^2) + g_M(k^2)}{2M} \tau^+_i \bm{\sigma}_i \delta(\bm{x} - \bm{r}_i).
\end{eqnarray}
Here $g_A(0)=1.27$, $g_V(0)+g_M(0) = 4.706$ and $M$ is mass of nucleon.
Within this approximation, the $<\lambda>$ term vanishes and 
only the $<\eta>$ term remains.
Therefore, the $0\nu\beta\beta$ decay
for $0^+-0^+$ transition
can be sensitive probe for the $<\eta >$ mechanism of $R$-handed current.
Assuming the s-wave electron wave function can be approximated by
constant, the amplitude $R_{0\nu}$  is given by the 
NME of two-body operator $M_{ij}$ as
\be
R_{0\nu}=4\sqrt{\frac{1}{2}}\left(\frac{G\cos\theta_c}{\sqrt{2}}\right)^2
\overline{e}_{p_2,s_2'}(\bm{0})\gamma_5 \gamma_0
e_{p_1,s_1'}^c(\bm{0})
         \braket{F|\sum_{i >j=1}^{A} M_{ij} |I},
\label{R0nu3}
\ee
with 
\begin{eqnarray}
  M_{ij} & = &    <\eta> \tau_i^+ \tau_j^+\int\frac{d\bm{k}}{(2\pi)^3}
  \frac{e^{i\bm{k}\cdot (\bm{r}_i-\bm{r}_j)}}{\omega}
  \left[ \frac{1}{\omega+A_1} + \frac{1}{\omega+A_2}\right]
   g_A(k^2)\frac{g_V(k^2) + g_M(k^2)}{2M} \nonumber \\
   & \times & [\frac{2}{3}\bm{k}^2 \bm{\sigma}_i\cdot \bm{\sigma}_j
     - (\bm{k}\cdot\bm{\sigma}_i  \bm{k}\cdot\bm{\sigma}_j - \frac{1}{3}
     \bm{k}^2 \bm{\sigma}_i\cdot \bm{\sigma}_j
     )
   ] .
\end{eqnarray}
The neutrino-exchange two-body operator $M_{ij}$, whose spin and momentum structure
is similar to the $\rho$ meson exchange nucleon-nucleon potential. After $\bm{k}$ integration,
effective two-body nuclear operator consists of spin-spin and tensor interaction.
  It is noticed that even though the light neutrino exchange mechanism,
  the operator becomes contact two-nucleon interaction
 by approximating $\omega \sim k$ and $A_i \sim 0$.
The short distance nature of the operator is well recognized and has been examined in
detail. It is essential to include both the finite size of
weak nucleon current for the effective nuclear operator and the short range correlation
for the nuclear wave function.
The form factors of axial vector and weak
magnetism of nucleon are usually parameterized in a dipole form
as $g_A(k^2) = g_A/(1 - k^2/1.14{\rm GeV}^2)^2$ and
$g_V(k^2) + g_M(k^2) \sim (g_V + g_M)/(1 - k^2/0.71{\rm GeV}^2)^2$.
Improved form factors and their uncertainties
are analyzed from the analyses of electron and neutrino scattering data on
proton and deuteron~\cite{ye2018,meyer2016}.
The short range correlation, which is not taken into account
in the model nuclear wave function, is taken into account by introducing short range
correlation (SRC) function \cite{M-S, Wu, Simkovic, Song}
\be
F(r)=1-ce^{-ar^2}(1-br^2)
\label{SRC1}
\ee
with $r=|\bm{r}_j-\bm{r}_i|$. 
This correlation function vanishes at $\lim r=0$ when $c=1$.  The suppression rates relative to those without the SRC are not affirmative, depending on nuclear models and various SRCs, from $5\%$ to $30-40\%$. See the most recent result, Fig.10 of  \cite{Yao}.

\begin{figure}

\includegraphics[width=10cm]{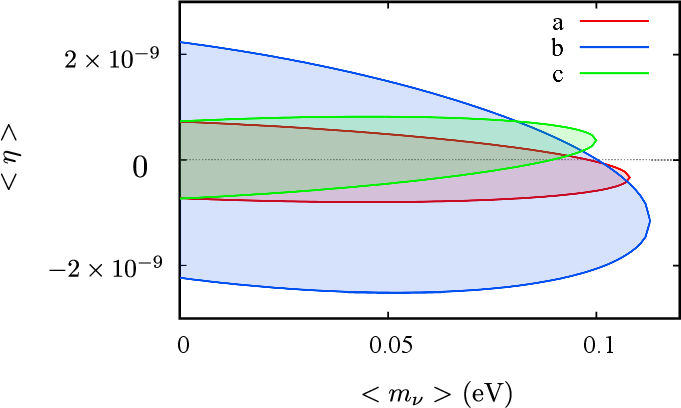}
 
\caption 
{ Allowed region of $<\eta>$ and $< m_\nu>$ for ${}^{136}Xe$.
  a,b,c are evaluated using $C$'s of Refs. \cite{muto89},\cite{suhonen91}
  and \cite{Pantis} (model without p-n pairing), respectively.
}
\label{fig:eta}
\end{figure}

The role of BSM physics on the 0$\nu\beta\beta$ decay
  has been studied extensively.(see references cited in \cite{agostini22}.)
Here we focus on $<\eta>$ and mass mechanisms.
Using the typical value of the ratio $C^{(0)}_{\eta\eta}/C^{(0)}_{mm} \sim 10^4$ to $10^5$,
the decay rate for $<m_\nu> \sim 100$ meV corresponds to $<\eta> \sim 10^{-9}$.
Actually, the decay rate is quadratic function of $<\eta>$ and $<m_\nu>$
for $<\lambda>=0$. 
Using current lower limit of $T_{1/2} > 2.3 \times 10^{26}$ years
from KamLAND-Zen \cite{KamLAND} and $C$'s given in Table 27 of
Ref. \cite{suhonen98},  $<\eta>$ and $<m_\nu>$ constrained from
the data are inside ellipses shown in Fig. \ref{fig:eta}.
Here not only $V-A$ interference terms ($\chi_R, \chi_P$),
$A-A$ and $V-V$ terms ($\chi_{1-}, \chi_{2+}$) are included.

The decay rate of a single nuclear species is not possible
to reveal the mechanism of $0\nu\beta\beta$ decay including BSM physics.
However, difference between
the space and spin structure of the
the effective nuclear operators for $<m_\nu>$ and $<\eta>$ mechanisms
has a potential to generate A-dependence of the decay rate.
A-dependence of the decay rate can be quantified by the
normalized decay rate with respect to the reference
 $0\nu\beta\beta$ process
for two extreme cases $<m_\nu>$ alone ($\alpha=m_\nu$) and
$<\eta>$ alone ($\alpha=\eta$) as
\begin{eqnarray}
  R_A^\alpha & = & \frac{T_{1/2}^\alpha(A_{\rm reference})}{T_{1/2}^\alpha(A)},
\end{eqnarray}
and deviation of the ratio $R_A = R_A^\eta/R_A^{m_\nu}$ from one
indicates an ability of the process to find BSM physics.
We estimated $R_A$ using $C$'s from the Tables of \cite{suhonen98}.
In table \ref{table2}, $R_A^\alpha$ and $R_A$ are calculated for
\cite{Pantis} (model without p-n pairing), where  ${}^{136}$Xe
is chosen as $A_{\rm reference}$. In this model, $R_A$ is appreciably small
for $^{48}$Ca, $^{82}$Se, $^{96}$Zr, $^{116}$Cd.
The result suggests that A-dependence of the decay rate can
be a key to disentangle the mechanism of 0$\nu\beta\beta$ decay and
to discover the BSM signal of $R$-handed current.
\begin{table}
\begin{center}
\begin{tabular}{c|cccccccc}\hline 
\hline
  Nucleus           & $^{48}$Ca & $^{76}$Ge & $^{82}$Se & $^{96}$Zr & $^{100}$Mo & $^{116}$Cd & $^{128}$Te & $^{130}$Te  \\ \hline
 $R_A^{m_\nu} $         & 0.75      &  0.51    & 1.2      & 3.0      & 0.47      &  0.39     &  0.095      &  2.1 \\
 $R_A^{\eta}$           &  0.082    & 0.40     &  0.19    & 0.83     &  0.36      &  0.064   &  0.10     & 2.0 \\ \hline 
 $R_A=R_A^{\eta}/R_A^{m_\nu}$ &  0.11    & 0.77     &  0.15    & 0.28     &  0.76      &  0.16   &  1.1     & 0.94 \\ \hline 
\end{tabular}
\end{center}
\caption{   Ratio of decay rate $R_A^\alpha$ evaluated using $C$'s of \cite{Pantis}.
} 
\label{table2}
\end{table}
 A precise A-dependence of $R_A$  may not yet established theoretically      
and it is expected to narrow down model dependence of NME,
especially for interaction range comparable with the nucleon size.

As for the heavy neutrino exchange mechanism,  there may be some enhancements
\cite{Cirigliano2, Li, Prezeau} due to
  $\pi\pi ee$ vertex from the effective operator, which generate pion range interaction,
\be
\mathcal{O}_{1+}^{ab}=\left(\overline{q}_L\tau^a\gamma^\mu q_L\right)\left(\overline{q}_R\tau^b\gamma_\mu q_R\right).
\ee
We can obtain the NME due to this term using the master formula by \cite{Cirigliano1,Cirigliano2} \footnote{We are grateful to Gang Li for useful discussions on this point.}
\be
\mathcal{M}_\nu^{(9)}=-\frac{1}{2m_N^2}C_{\pi\pi L}^{(9)}\left(\frac{1}{2}M_{GT}^{AP}+M_{GT}^{PP}+\frac{1}{2}M_T^{AP}+M_T^{PP}\right).
\ee
Here the notations and definitions are due to Appendix A.2 of \cite{Cirigliano2}. Thus, this term enhances $<\lambda>$ mechanism. However it may not change our order estimations because the original $C_{\eta\eta}$ is much larger than $C_{ij}$ with $i,j =m,\lambda$ by several orders.

\section{Conclusion}
We have studied $0\nu\beta\beta$ decay in the presence of $R$-handed current. We have tried to clarify the arguments of hadronic side and lepton's BSM physical one. As is well known, if neutrino masses obey IH, $<m_\nu>$ mechanism works around $50$ meV already marginal to the present and near future experiments. From NME, $<\lambda>$ mechanism is suppressed and $<\eta>$ mechanism can dominate the decay rate even around the present or near future experimental limits. This is the case if the neutrino masses belong to NH which is much more probable than IH from theoretical reasons. Even if we get the non-null result in $0\nu\beta\beta$ decay in a single species, though it is the great achievement, we can not limit BSM physics. It is very important to survey
this process in different nuclei to specify BSM physics.

\section*{Acknowledgments}
We would express our sincere thanks to Dr. H. Ejiri for his encouragements.
This work is supported by JSPS KAKENHI Grant Numbers JP22H01237 (T.F) and JP19H05104 (T.S).

\end{document}